\documentclass[pdflatex,sn-basic]{sn-jnl}

\usepackage{graphicx}
\usepackage{amsmath,amssymb,amsfonts}
\usepackage{longtable}
\usepackage{booktabs}
\usepackage[title]{appendix}
\usepackage{placeins}

\hypersetup{
  pdftitle={Integrating Flipped Learning and Generative AI for Practice-Based Design Education: Evidence from a Knit Yarn Design Course},
  pdfauthor={Hong Qu; Zichao Ling; Yadie Yang},
  pdfsubject={Flipped learning and generative AI in practice-based design education},
  pdfkeywords={Flipped learning, Generative AI, Design education, Visual prototyping, Knit yarn design}
}

\begin{document}

\title[Integrating Flipped Learning and Generative AI]{Integrating Flipped Learning and Generative AI for Practice-Based Design Education: Evidence from a Knit Yarn Design Course}

\author[1]{\fnm{Hong} \sur{Qu}}

\author[1]{\fnm{Zichao} \sur{Ling}}

\author*[1]{\fnm{Yadie} \sur{Yang}}

\affil[1]{\orgdiv{School of Fashion and Textiles}, \orgname{The Hong Kong Polytechnic University}, \orgaddress{\city{Hong Kong}, \country{China}}}

\abstract{In practice-based design courses such as knit yarn design, students must turn visual ideas into feasible material outcomes. This is difficult because creative decisions are tied to yarn properties, stitch structures, machine operation, and limited opportunities for physical sampling. This study presents an integrated pedagogical framework that combines flipped learning, exemplar-based reference, GenAI-assisted visual prototyping, and studio feedback in an undergraduate knit yarn design course. The framework was implemented through a cross-device platform with pre-class micro-videos, formative checks, a curated gallery, and a GenAI-supported ideation module. An exploratory course-based evaluation compared a historical control cohort (\(N=12\)) and an intervention cohort (\(N=16\)), supplemented by questionnaire responses and brief interviews. The findings are interpreted as context-specific indicators rather than confirmatory causal evidence. Exploratory comparisons showed higher scores in creativity thinking, design skills, problem solving, and total course score in the intervention cohort. Student and instructor responses suggested that flipped learning supported studio readiness, while GenAI mainly supported early-stage visual exploration rather than precise technical guidance. Overall, the study offers a practice-based instructional framework for integrating flipped preparation, GenAI-assisted visual prototyping, and studio feedback in design education.}

\keywords{Practice-based studio courses , Studio-based learning , Generative artificial intelligence (GenAI) , Flipped learning , Visual prototyping , Knit yarn design}

\makeatletter
\g@addto@macro\auaddress{\par\small *Corresponding author: Yadie Yang.}
\makeatother
\maketitle
\section{Introduction}
\label{intro}

Practice-based design education develops knowledge through making, testing, feedback, and revision \citep{practice-based-design-education2009, practice-design-education2020}. Students are expected to externalize ideas, work with materials and tools, respond to constraints, and refine artifacts over time \citep{iteration-design2011,Chimwayange2025PBLServiceSkills}. Studio-based learning is a common form of this tradition because it centers making, critique, and iterative improvement \citep{cennamo2011managing,Kumar2021StudioBasedLearning}. In this study, studio-based learning is treated as the pedagogical context in which the course intervention was implemented.

A central challenge is that design learning depends on repeated experimentation and feedback. Students need to generate ideas, make design decisions visible, receive critique, and revise their work across multiple cycles \citep{cennamo2011managing,Kumar2021StudioBasedLearning}. Yet this process is shaped by time, materials, equipment access, technical support, and instructor availability \citep{material-constraints-design2010,practice-design-education2020}. When iteration is costly, students may explore fewer alternatives or settle too early on safer solutions. This weakens the feedback--revision loop that supports practice-based learning \citep{dow2009}, especially in courses that require repeated physical trials \citep{Noguchi1999}.

A second challenge concerns preparation. Studio sessions are most valuable when they focus on demonstration, hands-on practice, critique, coaching, and the refinement of work-in-progress \citep{Megahed2018,Kumar2021StudioBasedLearning}. However, students often arrive with uneven procedural knowledge, technical vocabulary, and understanding of assessment expectations \citep{practice-based-design-education2009}. Class time may then be spent on repeated explanation rather than making and feedback \citep{AKCAYIR2018334, Shibukawa2019}. This matters in material-oriented design courses, where weak preparation can delay experimentation and reduce the quality of in-class critique \citep{knitwear-design2017,material-constraints-design2010}.

Flipped learning can address this problem by moving selected foundational instruction before class and reserving contact time for active engagement, guided practice, and problem solving \citep{AKCAYIR2018334, Shibukawa2019}. Prior work suggests that its effectiveness depends on engagement, preparation, and self-regulated learning \citep{chen2018academic,geng2025}. Pre-class activities therefore need to be structured and connected to in-class work, rather than treated as optional content delivery \citep{rasheed2020self}. In design courses, such preparation should help students understand constraints, make decisions, and participate more productively in critique and revision \citep{Kumar2021StudioBasedLearning}.

Preparation also needs to be accessible. If pre-class work depends on specific devices, software, locations, or studio access, students may be less likely to complete it \citep{mobile-education2020,cross-device-learning2025}. Cross-device platforms can reduce this friction by allowing students to review videos, complete checks, consult exemplars, and prepare ideas across different learning situations \citep{cross-device-learning2025}. This is useful in design courses, where technology-enhanced studio practices increasingly support learning beyond scheduled class time \citep{SunerPlaCerda2025StudioTechnologyReview}.

Generative artificial intelligence (GenAI) may also reduce the cost of early-stage iteration. Image-generation systems can produce diverse visual alternatives and can therefore act as low-cost visual prototyping tools \citep{zhang2024,Smith2025}. They can support early exploration, comparison, and critique before students commit time and materials to physical sampling \citep{ray2024,pires2025prompts,qu2025recycling}. This potential is relevant to design courses in which students must balance creative exploration with limited materials, equipment access, and technical feasibility \citep{juan2026,yang2025}. GenAI can therefore support ideation before making, without replacing hands-on practice \citep{yang2025,Chen2026GenAIMetacognitionDesign}.

The educational value of GenAI, however, cannot be assumed. It may encourage shallow trial-and-error behavior, over-reliance on tool outputs, weak authorship, or poor alignment with domain constraints \citep{Reynolds2024,genai-design2026cognitive}. These risks matter in design education because successful work depends not only on visual appeal, but also on material logic, technical feasibility, process understanding, and justified design decisions \citep{material-constraints-design2010,urcola2024design}. GenAI-supported visual prototyping therefore needs to be connected to clear criteria, reflective comparison, and subsequent physical making \citep{genai-learning2025scaffolding}.

Despite growing interest in GenAI in education, evidence remains uneven. Many studies focus on text-centered tasks and outcomes such as writing performance, conceptual tests, participation, and self-reported experience \citep{LAW2024100174,Weng2025,CHAPELLE2025103672}. Less is known about GenAI-supported learning in artifact-centered design courses, where evaluation must consider both creative intention and practical feasibility \citep{PERDANA2026102035,ray2024}. This gap matters because early visual exploration may not improve design outcomes unless students are also supported in preparation, comparison, critique, and physical realization \citep{zhang2024,Megahed2018}.

Knit yarn design is a useful context for examining this issue because it is both creative and technically constrained. Students must coordinate stitch structures, yarn properties, colour decisions, knitting techniques, visual composition, and repeated physical sampling \citep{knit_design2011,knitwear_design_case2004}. They must judge whether an intended visual effect can become a feasible knitted artifact \citep{knitwear_design_case2004}. Because physical sampling can be time-consuming and resource-intensive, early-stage visual prototyping is especially relevant \citep{Noguchi1999}.

This study addresses the gap by proposing and examining a pedagogical framework for integrating flipped learning and GenAI-assisted visual prototyping in an undergraduate knit yarn design course. The course used a cross-device platform that connected preparation, exemplar review, AI-supported ideation, and studio-based refinement. The platform included pre-class videos, embedded checks, and a curated gallery of prior student work \citep{geng2025,LO201750}. It also included a GenAI module based on a vision-language model. GenAI was used to support ideation, not to produce final assessed artifacts. Students could generate and compare alternatives before selecting directions for physical realization. To reduce prompt-writing barriers, the module provided optional templates and parameterized options for specifying key design dimensions \citep{juan2026,juan2026scaffolded}.

The platform also enabled an exploratory course-based evaluation. Learning traces, including video engagement, embedded checks, platform use, and GenAI generation frequency, were examined alongside achievement \citep{BARNARD20091,chen2018academic}. The study compared two cohorts taught by the same instructor and assessed with the same rubric: a prior cohort without the platform-supported intervention and an intervention cohort with the flipped-learning, cross-device, and GenAI-supported design. Because the comparison was historical rather than randomized, the quantitative findings are interpreted as exploratory indicators rather than causal evidence. Across cohorts, the assessed components and their weighting were held constant. Major deliverables were graded using analytic rubrics with component-level criteria \citep{eshun2013design,Kumar2021StudioBasedLearning}.

The contribution of the study is primarily pedagogical. First, it articulates a framework for organizing preparation, exemplar reference, AI-assisted visual exploration, and material refinement in a design course. Second, it shows how a cross-device platform can connect preparation, classroom activity, feedback, and revision. Third, it positions GenAI-supported visual prototyping as a scaffold for early exploration rather than a replacement for hands-on making or instructor judgement. Although the empirical focus is knit yarn design, the framework may inform other design courses that combine visual ideation, material experimentation, and technical development.

\section{Platform Implementation}
\label{platform}

\begin{figure}[htbp]
  \centering
  \includegraphics[width=\linewidth]{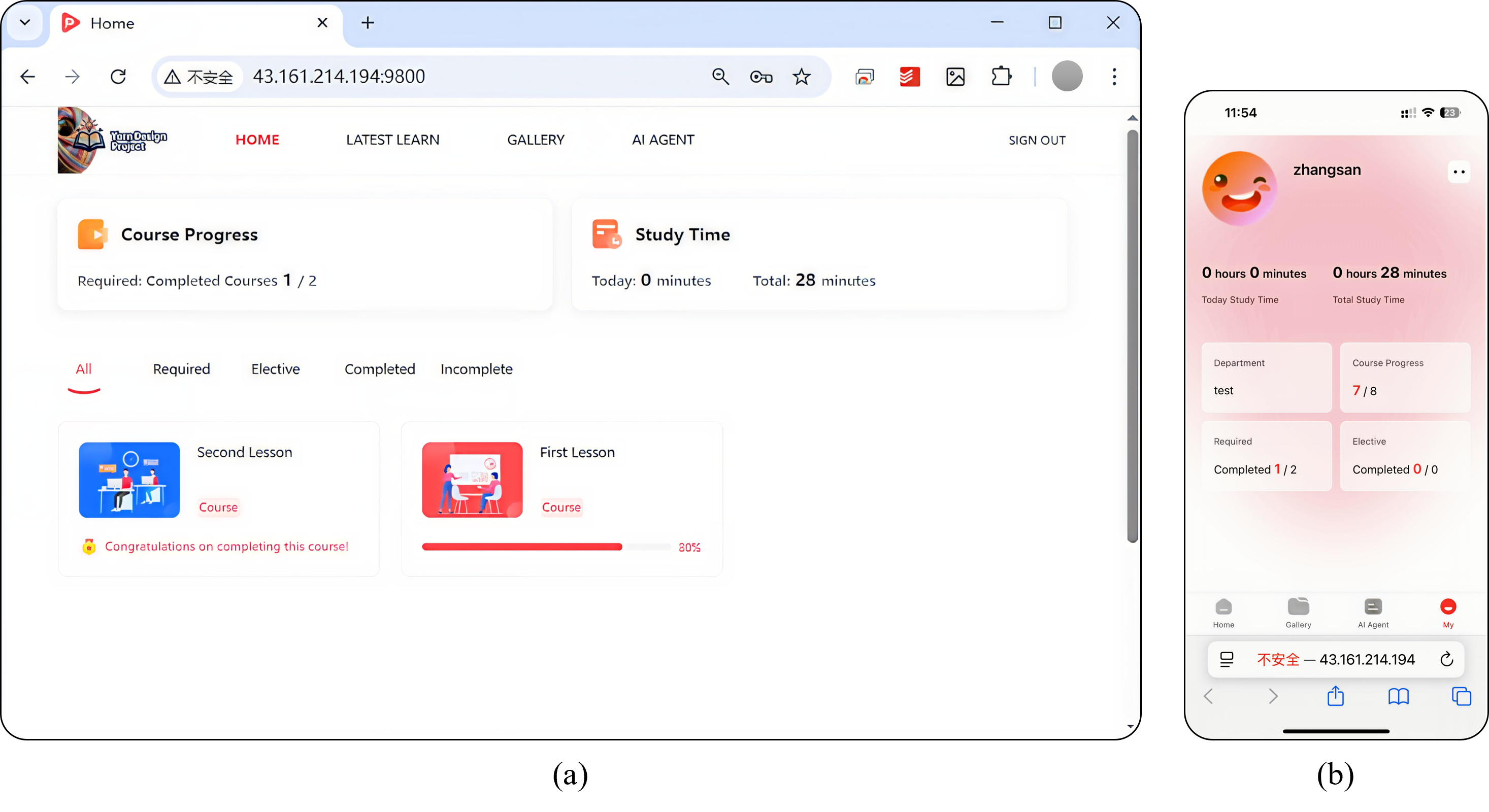}
  \caption{Cross-device platform interface. (a) Desktop interface showing two lesson modules and learning progress indicators. (b) Mobile interface showing the responsive layout of the platform.}
  \label{fig:website}
\end{figure}

We developed a cross-device learning platform as a responsive website to support pre-class learning, studio readiness, and early-stage visual ideation in an undergraduate knit yarn design course (Fig.~\ref{fig:website}). Students could access course materials, readiness checks, gallery examples, and GenAI-supported ideation tools on desktop and mobile devices. The platform was intended to reduce the practical burden of preparation and support continuity between preparation, classroom activity, feedback, and later sampling.

In line with flipped learning principles, the platform provided structured pre-class materials and formative readiness checks. The course consisted of four lecture sessions followed by nine studio sessions; the final studio session was used for the final presentation. The platform complemented the scheduled lectures and helped students prepare for later studio operations, critique, and physical sampling.

The platform was designed not only as a teaching tool but also as an instrumented learning environment. Students' preparation and exploration behaviours could therefore be captured during routine course participation and examined alongside learning outcomes. The platform comprised three main components: a Course module, a Gallery module, and a GenAI module, which was labeled ``AI Agent'' in the platform interface.

\paragraph{Course module}

The lecture sessions introduced foundational knowledge related to yarn production, knitting processes, machine operation, yarn selection, and design considerations. The studio sessions focused on hands-on sampling, material experimentation, design development, and portfolio preparation. The Course module supported the early lecture-based phase through short video lessons and embedded readiness checks. It helped students review selected content before later lecture and studio activities, rather than replacing teacher-led instruction.

Content was organized into two lessons comprising eight micro-videos in total, each approximately five minutes in length. These lessons were aligned with the early lecture phase of the course. Lesson~1 was assigned after the first lecture as preparation for the following lecture, and Lesson~2 was assigned after the second lecture as preparation for subsequent lecture and studio activities. Thus, the Course module supported students' first exposure to and review of selected foundational content outside scheduled class time, while the four lecture sessions continued to provide formal instruction, explanation, clarification, and connection to later studio work.

\begin{figure}[htbp]
  \centering
  \includegraphics[width=\linewidth]{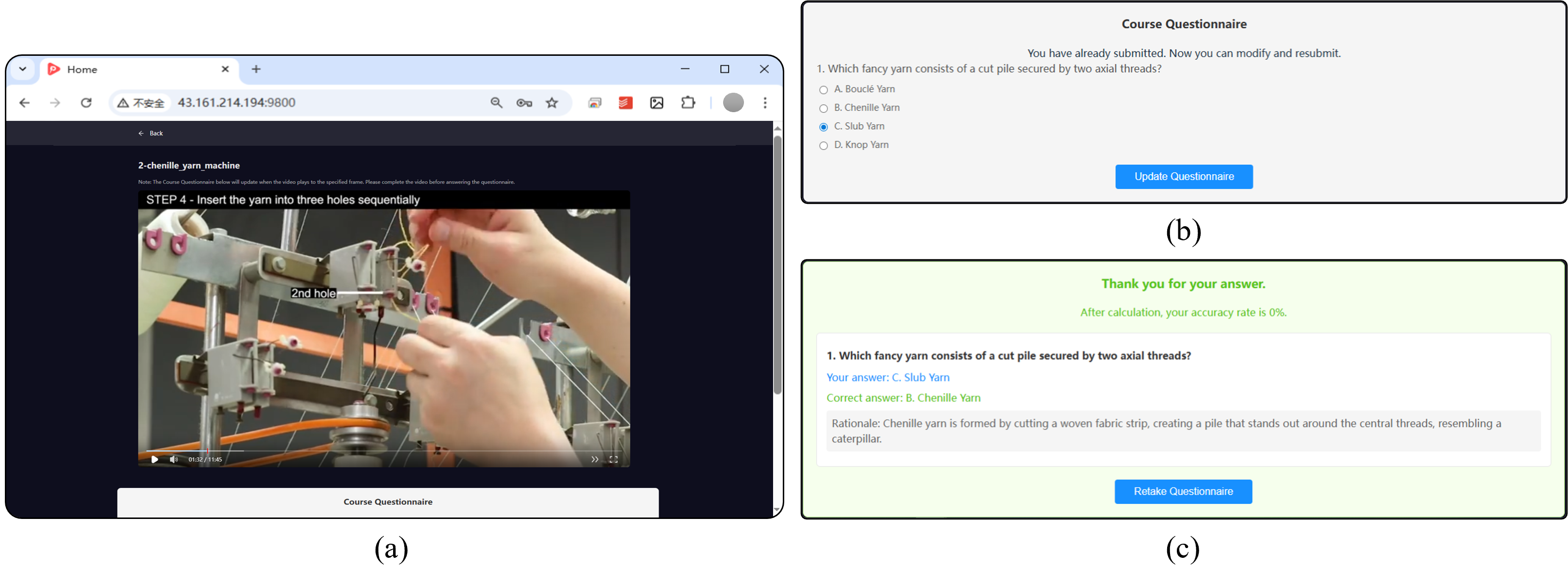}
  \caption{Course module workflow for flipped pre-class learning and embedded readiness checks. (a) Pre-class micro-video lessons with progress indicators, (b) embedded single-choice and multiple-choice readiness checks, and (c) post-submission feedback with correctness marking and explanations.}
  \label{fig:video_lesson}
\end{figure}

The interface displayed progress states, such as not started, in progress, and completed, to help students monitor their preparation (Fig.~\ref{fig:video_lesson}(a)). Each video segment was paired with embedded readiness checks, including both single-answer and multiple-answer items (Fig.~\ref{fig:video_lesson}(b)). Students could retry the items, and the system provided correctness feedback together with brief explanations after submission (Fig.~\ref{fig:video_lesson}(c)). These readiness checks were formative rather than summative. They were intended to reinforce key concepts, make preparation visible, and help students identify gaps before subsequent lecture and studio sessions.

To align preparation with engagement, each readiness check was unlocked only after the corresponding video segment had reached a predefined viewing threshold. Interface functions that would have allowed students to bypass this requirement were disabled. This design was intended to encourage engagement with the pre-class learning materials before students attempted the embedded checks.

In the fourth lecture session, students participated in a student learning presentation activity in which they explained and discussed what they had learned from the lectures and the platform-supported pre-class preparation. This activity was used to consolidate foundational understanding, make students' preparation visible, and connect early lecture-based learning with later studio-based design work.

\paragraph{Gallery module}

\begin{figure}[htbp]
  \centering
  \includegraphics[width=\linewidth]{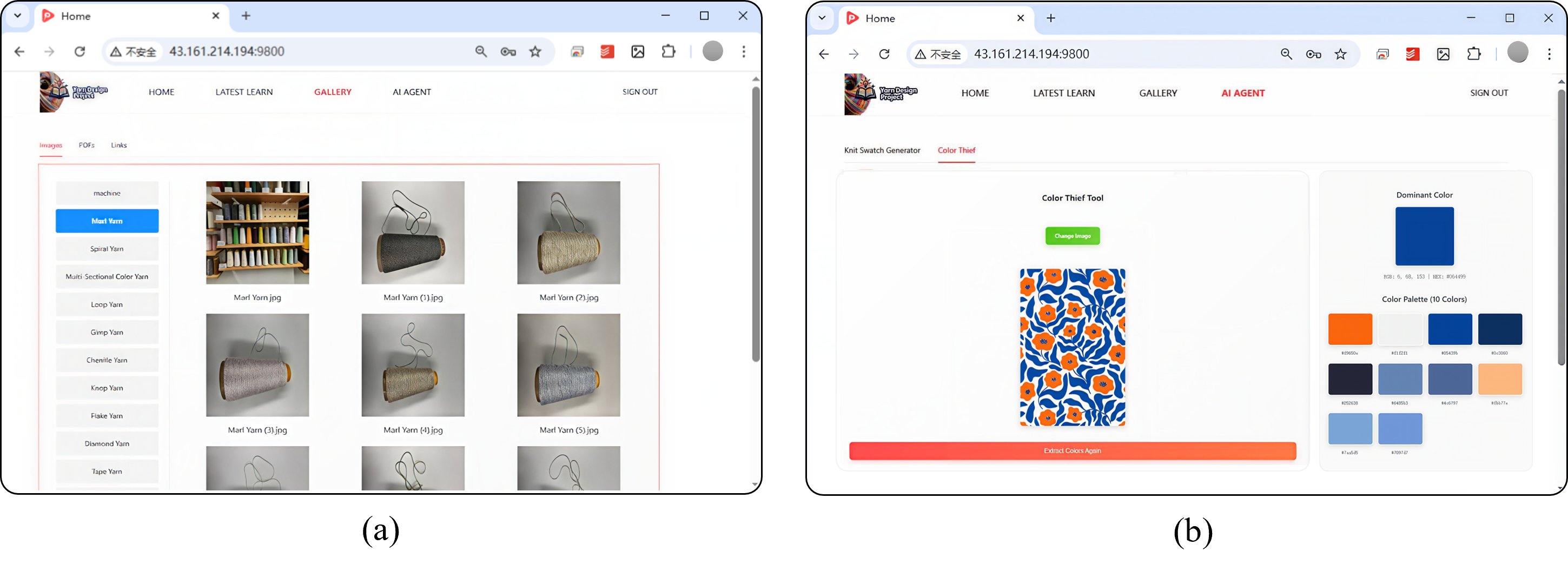}
  \caption{Platform support for exemplar browsing and color exploration. (a) The Gallery module presents yarn-design exemplars organized by yarn type. (b) The Color Thief palette tool generates a 10-color palette and corresponding hexadecimal color codes from an uploaded image.}
  \label{fig:gallery_color}
\end{figure}

The Gallery module provided curated exemplars to help students understand feasible outcomes under workshop and material constraints and to calibrate quality expectations. The gallery was organized by yarn type and covered thirteen categories that could be produced in the lab, such as tape yarn, chenille yarn, and slub yarn (Fig.~\ref{fig:gallery_color}(a)). In addition to exemplar images, the gallery included selected high-quality work from prior cohorts and supplementary links for extended reading and self-study.

To preserve the validity of the cohort comparison, the gallery did not include student work from the historical control cohort used in this study. Instead, the exemplar set was drawn from earlier teaching materials and prior cohorts outside the two cohorts included in the outcome comparison. This design choice was intended to prevent the intervention cohort from being exposed to the historical control cohort's assessed artifacts while still providing domain-relevant exemplars for quality calibration.

\paragraph{GenAI module}

\begin{figure}[htbp]
  \centering
  \includegraphics[width=0.8\linewidth]{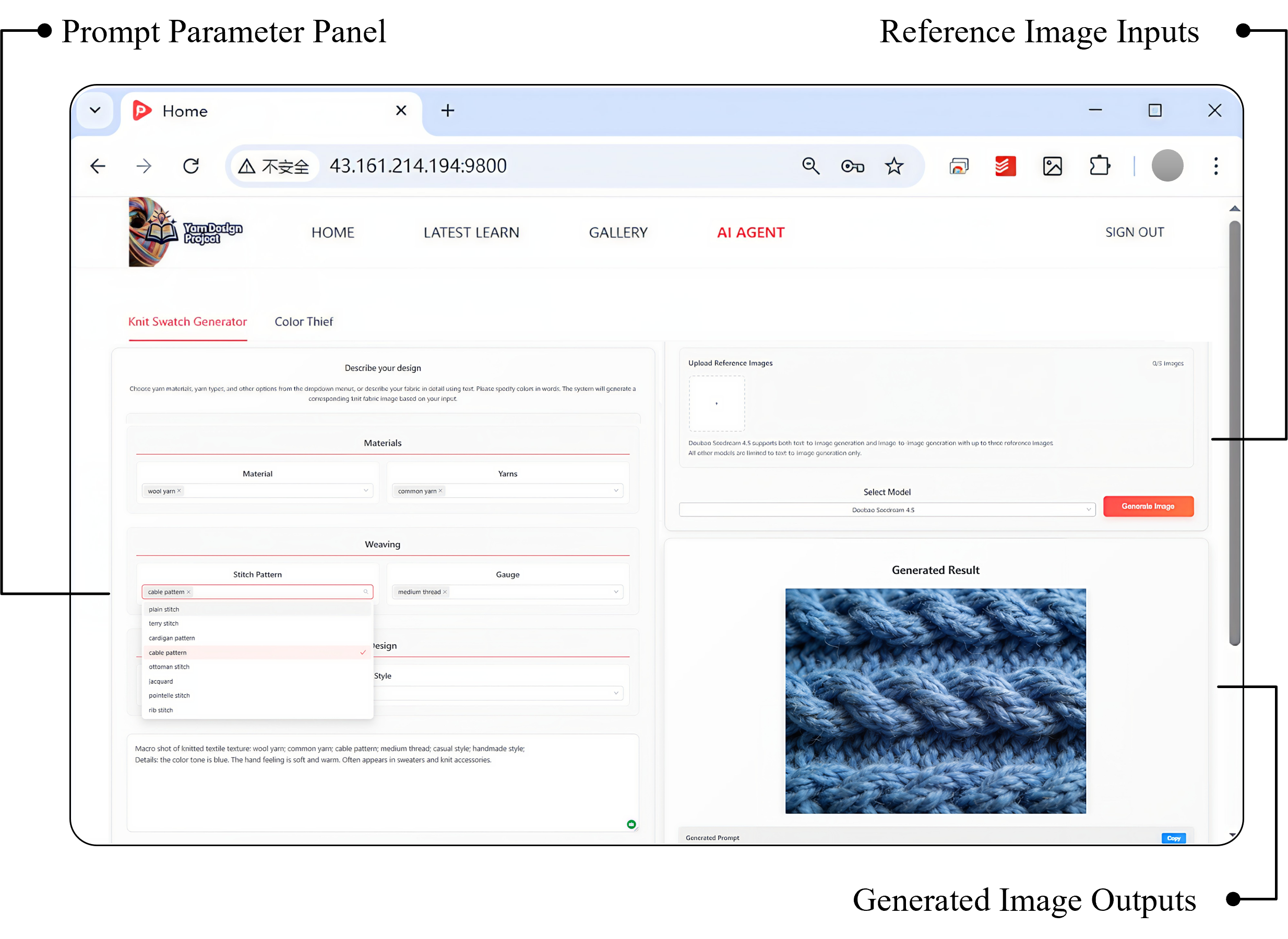}
  \caption{Parameter-guided prompt scaffolding interface in the GenAI module. The interface supports scaffolded prompt construction through a prompt parameter panel and optional reference image inputs, and presents generated visual outputs for early-stage design ideation.}
  \label{fig:genai_interface}
\end{figure}

The GenAI module supported early-stage ideation by interfacing with a commercial image generation model through an application programming interface (API)~\footnote{https://seed.bytedance.com/zh/seedream4\_5}. To reduce prompt-writing barriers and support more domain-relevant exploration, the module provided a parameter-guided prompt scaffolding interface (Fig.~\ref{fig:genai_interface}). Students could construct prompts by selecting and combining design parameters, including color direction or palette, yarn type, knit or stitch structure, texture, material effect, and visual style. They could also manually revise or extend the generated prompts for greater control and, when appropriate, provide optional reference images to guide visual generation.

\begin{figure}[htbp]
  \centering
  \includegraphics[width=\linewidth]{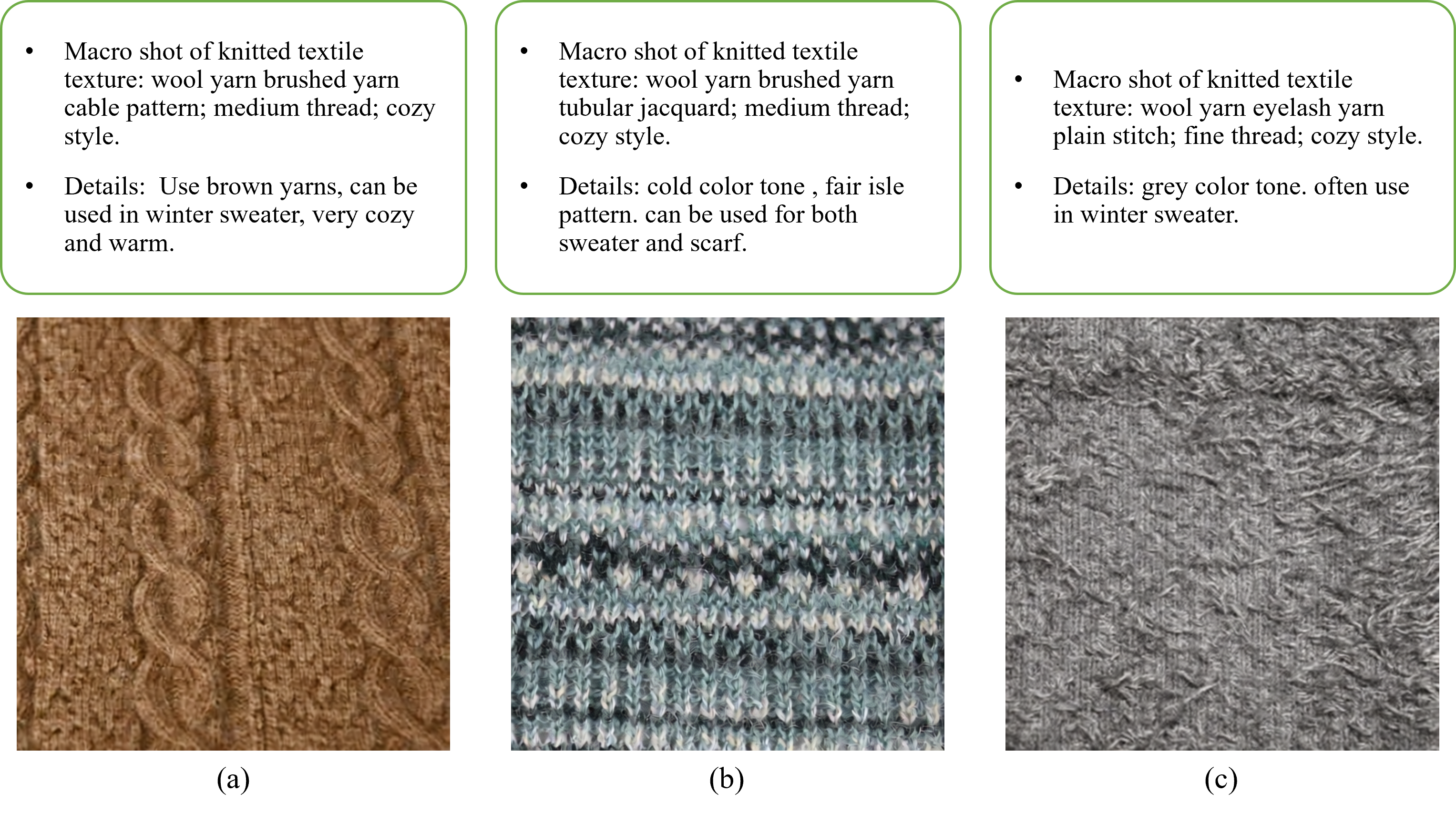}
  \caption{Example outputs from the GenAI-assisted visual prototyping workflow. The images demonstrate how different parameter combinations could produce varied visual references in texture, stitch expression, color direction, and surface composition. These outputs were used as ideation prompts rather than as technically accurate or directly manufacturable textile designs.}
  \label{fig:genai_results}
\end{figure}

These parameter categories were derived from practical dimensions of knit yarn and knitted fabric design rather than generic image-generation keywords. As illustrated in Fig.~\ref{fig:genai_results}, students could specify a color direction, select a yarn category, indicate a stitch or structural tendency, and combine these choices with surface qualities and visual styles. The interface helped students express design intentions using terms close to the material, structural, and visual concerns of the course. Students could also upload one to three reference images as auxiliary visual inputs.

The generated images functioned as visual references for ideation, comparison, and critique, not as simulations of final textile outcomes. Current text-to-image models have limited understanding of yarn behaviour, stitch formation, knit structure, material properties, and workshop feasibility. Generated images could suggest texture, color mood, surface composition, or style combinations, but they could not reliably represent executable knitting instructions or physically feasible textile outcomes. Students were therefore instructed to treat the outputs as early-stage prompts that required interpretation and later hands-on testing.

Students could generate alternatives, compare directions, and download outputs before committing to physical sampling. The module therefore served as a low-cost visual prototyping environment during concept development. Fig.~\ref{fig:genai_results} illustrates this workflow through example outputs that vary in texture, stitch expression, color direction, and surface composition.

To support color exploration, the module also incorporated a palette extraction tool based on Color Thief~\footnote{https://lokeshdhakar.com/projects/color-thief}. Students could upload an input image and obtain a 10-color palette together with hexadecimal color codes (Fig.~\ref{fig:gallery_color}(b)). These color codes could then be used to inform prompt construction, compare generated alternatives, or support later yarn and material selection.

\paragraph{Instrumentation and learning traces}
To support the measurement-forward design of the study, the platform logged learning traces across modules. In the Course module, logs included video watch time, completion status, and readiness-check records. In the Gallery module, logs captured visit frequency. In the GenAI module, logs included image generation frequency, parameter selections in the scaffolded interface, and related interaction records. These traces were later used as behavioral indicators of preparation and exploration in the analysis.

\section{Methodology}
\label{methodology}

\begin{figure}[htbp]
  \centering
  \includegraphics[width=\linewidth]{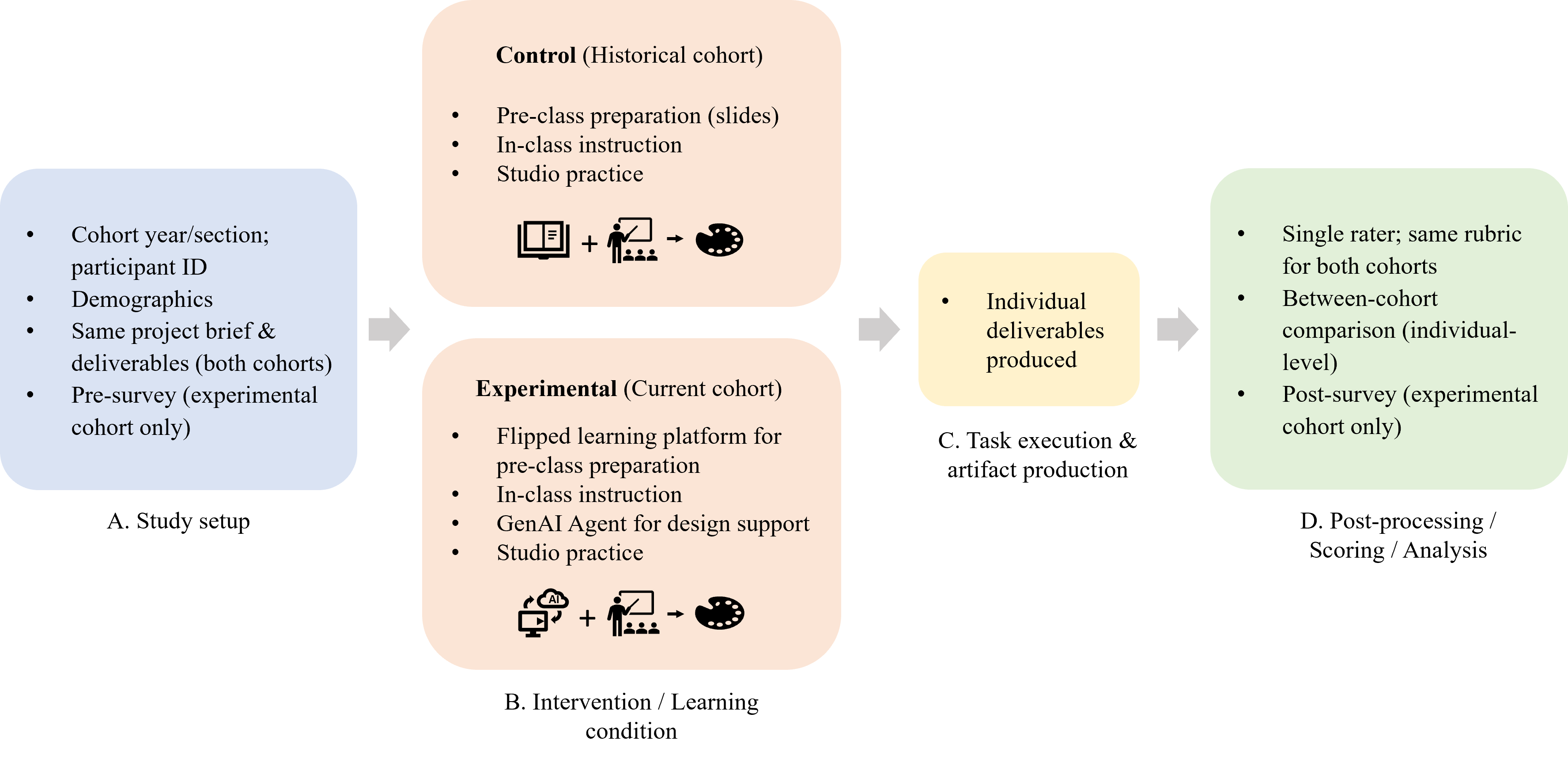}
  \caption{Overview of the study design and data collection procedure.}
  \label{fig:method}
\end{figure}

\subsection{Study design}

This study adopted a quasi-experimental historical-cohort design to examine whether a platform-supported flipped-learning intervention with integrated GenAI-assisted ideation was associated with differences in learning outcomes in a knit yarn design studio course. The historical control cohort was the previous-year offering of the course (\(n=12\)), and the intervention cohort was the current-year offering (\(n=16\)). The small cohort size reflects the authentic organization of practice-based studio teaching, where intensive demonstration, machine access, critique, and iterative feedback require limited enrolment. Accordingly, the study does not aim to provide large-scale generalizable evidence, but to examine how an integrated flipped-learning and GenAI-supported design can operate within a materially constrained studio course. Figure~\ref{fig:method} summarizes the overall study procedure.

Both cohorts consisted of third-year undergraduate students enrolled in the same course taught by the same instructor. The two cohorts followed the same project brief, major deliverables, assessment rubric, and grading scheme. The main difference between the two course offerings was that the intervention cohort used the platform-supported flipped-learning arrangement and the GenAI-assisted ideation module, whereas the historical control cohort followed the earlier version of the course.

Because students were not randomly assigned to cohorts and the comparison was made across two academic years, the design cannot fully rule out cohort differences or other contextual factors. The study was therefore positioned as an exploratory evaluation of an authentic course implementation. To support cautious interpretation, the analysis reports descriptive statistics, effect sizes, confidence intervals (CIs), and non-parametric sensitivity tests where appropriate.

\subsection{Course context, rubric, and learning outcomes}

The course followed a studio-based learning format focused on knit yarn and knitted fabric design. Student performance was assessed using a common rubric-based grading scheme applied across both cohorts. The rubric consisted of five dimensions: Creativity thinking, Design skills, Problem solving, Presentation, and Attendance. Each dimension was scored on a 0--100 scale, and the weighted total course score was calculated from these five dimensions.

The rubric dimensions and their weightings are shown in Table~\ref{tab:rubric_weights}. The same instructor assigned the course scores for both cohorts using the same rubric and weighting scheme.

\begin{table}[htbp]
  \centering
  \caption{Assessment rubric dimensions and weightings}
  \label{tab:rubric_weights}
  \begin{tabular}{p{0.62\linewidth} p{0.18\linewidth}}
    \hline
    \textbf{Dimension} & \textbf{Weight} \\
    \hline
    Creativity thinking & 35\% \\
    Design skills & 20\% \\
    Problem solving & 20\% \\
    Presentation & 15\% \\
    Attendance & 10\% \\
    \hline
  \end{tabular}
\end{table}

\subsection{Intervention}

\paragraph{Pre-class learning arrangement}

Both cohorts were expected to engage in pre-class preparation, but the form of preparation differed between them. In the historical control cohort, preparation primarily involved previewing slide materials before class. In the intervention cohort, preparation was supported through the cross-device platform, which provided structured pre-class videos, learning materials, examples, and formative checks. These materials were intended to help students become familiar with yarn production processes, machine operation, yarn selection logic, and relevant knitting procedures before entering the studio.

The flipped-learning arrangement was designed to move part of the basic procedural and conceptual preparation outside the studio session, so that in-class time could be used more directly for demonstration, hands-on practice, discussion, experimentation, and feedback.

\paragraph{GenAI-assisted ideation module}

In addition to the structured platform-based preparation, the intervention cohort had access to the GenAI-assisted visual ideation module described in Section~\ref{platform}. This module was intended to support early-stage exploration during the yarn design process, including ideation, comparison of alternative visual directions, and reflection on possible design variations.

The module was used as a support tool rather than as a replacement for instructor feedback, hands-on demonstration, or technical decision-making. In the course setting, its intended role was to help students generate and compare visual references during the creative process.

\subsection{Data collection}

\paragraph{Course scores}

For both the historical control cohort and the intervention cohort, the instructor-assigned score for each rubric dimension was collected, together with the resulting weighted total course score. All score outcomes were recorded on a 0--100 scale. These course scores formed the basis of the between-cohort comparison reported in Section~\ref{results}.

\paragraph{Questionnaires}

Questionnaire data were collected from the intervention cohort only. Students completed a pre-course questionnaire and a post-course questionnaire. Both questionnaires used 5-point Likert-type items, together with optional open-ended questions.

The questionnaire design was informed by previously published instruments and related studies on self-directed learning, motivation, design learning, and AI-supported educational experience \citep{BARNARD20091,urcola2024design,lee2024development,dongjiao2025research}. The pre-course questionnaire asked students to report their perceived baseline learning dispositions and abilities before participating in the intervention. The post-course questionnaire asked students to evaluate their learning experience after the course, including the flipped-learning component, the GenAI module, and the overall learning setting.

For the pre--post comparison, two conceptually aligned constructs were used: \textit{Innovation and Motivation} and \textit{Self-Directed Learning Ability}. Each construct was represented by four items in the pre-course questionnaire and four related items in the post-course questionnaire. However, the pre- and post-course items were not worded as identical repeated measures. Therefore, these comparisons were interpreted as shifts in self-reported perceptions from baseline to post-course evaluation, rather than as strict psychometric measures of latent construct growth.

The post-course questionnaire also included additional evaluative items on the flipped-learning component, the GenAI module, and the overall class setting. These items were analyzed descriptively to summarize students' perceptions of different aspects of the intervention. The full questionnaire instruments and item wording are provided in Appendix~A, Tables~\ref{tab:pre_survey} and~\ref{tab:post_survey}.

\paragraph{Brief interviews}

To provide contextual information for interpreting the course scores and questionnaire results, brief interviews were conducted with four students from the intervention cohort and the course instructor. The interviews focused on students' experiences with pre-class preparation, studio learning, GenAI-supported ideation, and the overall course setting.

The interviews were audio-recorded with participants' consent, transcribed, reviewed, and summarized. The interview data were used to provide illustrative explanations and contextual examples for patterns observed in the quantitative results.

\subsection{Data analysis}

Data analyses were conducted at the individual level. Continuous course-score outcomes were summarized using means and standard deviations. Questionnaire responses were summarized using construct-level means, standard deviations, favorable response rates, and item-level distributions where appropriate. Favorable responses were defined as ratings of 4 or 5 on the 5-point Likert scale.

\paragraph{Between-cohort comparison of course performance}

Between-cohort differences between the historical control cohort and the intervention cohort were examined for the weighted total course score and for each of the five rubric dimensions. Welch's two-sample \(t\)-tests were used because they do not assume equal variances between groups. Mean differences were calculated as intervention cohort minus historical control cohort.

Standardized mean differences were reported using Hedges' \(g\). Bootstrap 95\% CIs were estimated for the mean differences using percentile bootstrap resampling with individuals resampled within each cohort over 5{,}000 iterations. Mann--Whitney \(U\) tests were also conducted as non-parametric sensitivity analyses for the same between-cohort comparisons.

\paragraph{Pre--post questionnaire comparison}

For the intervention cohort questionnaire data, construct scores for \textit{Innovation and Motivation} and \textit{Self-Directed Learning Ability} were calculated by averaging the four items within each construct. Pre--post differences were calculated as post-course construct score minus pre-course construct score.

Because the pre- and post-course items were conceptually aligned but not identically worded, the pre--post results were treated as exploratory indicators of changes in self-reported perceptions. Wilcoxon signed-rank tests were used to examine pre--post differences for the two constructs. Bootstrap 95\% CIs were calculated for the mean pre--post changes. Favorable response rates were also reported to show the proportion of ratings at 4 or 5 before and after the course.

Post-course ratings of the flipped-learning component, the GenAI module, and the overall learning setting were analyzed descriptively. These results were used to identify which aspects of the learning setting students rated more or less positively.

\paragraph{Interview and open-ended response analysis}

Interview transcripts and open-ended questionnaire responses were reviewed to identify recurring points related to students' learning experiences. The review focused on comments about pre-class preparation, readiness for hands-on studio work, experimentation and refinement, portfolio development, and the perceived role and limitations of the GenAI module.

Because the qualitative dataset was small, the analysis was descriptive and interpretive rather than a formal attempt to generate generalizable qualitative findings. Representative quotations were selected to illustrate recurring observations and to help explain the quantitative patterns reported in the Results section. Claims based on these materials were therefore kept cautious and contextual.

\section{Results}
\label{results}

Given the historical-cohort design and small sample size, the quantitative findings are interpreted as exploratory indicators from an authentic course implementation rather than as confirmatory evidence of causal effectiveness. The results are therefore presented alongside questionnaire, interview, and illustrative case evidence to clarify how different components of the learning design were experienced and used.

\subsection{Between-cohort comparison of course performance}

\begin{figure}[htbp]
  \centering
  \includegraphics[width=\linewidth]{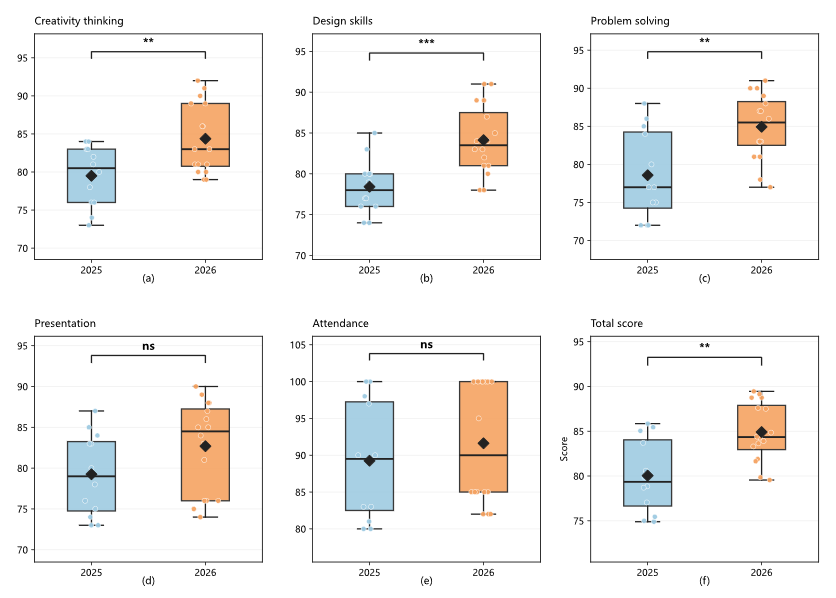}
  \caption{Between-cohort comparison of rubric dimension scores and total course score for the historical control cohort and the intervention cohort: (a) Creativity thinking, (b) Design skills, (c) Problem solving, (d) Presentation, (e) Attendance, and (f) total course score. Diamonds indicate group means. Welch's \(t\)-test results and Hedges' \(g\) are annotated in each panel. Asterisks indicate significance levels (* \(p<0.05\), ** \(p<0.01\), *** \(p<0.001\)).}
  \label{fig:score_comparison}
\end{figure}

\begin{table}[htbp]
  \centering
  \caption{Between-cohort comparison of course performance}
  \label{tab:between_cohort_performance}
  
  \footnotesize
  \setlength{\tabcolsep}{3.0pt}
  \begin{tabular}{lcccccc}
    \toprule
    Outcome & Hist. M (SD) & Int. M (SD) & Diff. [95\% CI] & Welch \(p\) & \(g\) & MW \(p\) \\
    \midrule
    Creativity thinking & 79.50 (3.97) & 84.38 (4.59) & 4.88 [1.81, 8.02] & 0.006 & 1.09 & 0.032 \\
    Design skills & 78.42 (3.40) & 84.13 (4.24) & 5.71 [2.98, 8.35] & \(<0.001\) & 1.42 & 0.001 \\
    Problem solving & 78.58 (5.85) & 84.94 (4.30) & 6.36 [2.40, 10.00] & 0.005 & 1.23 & 0.007 \\
    Presentation & 79.25 (5.05) & 82.69 (5.56) & 3.44 [-0.29, 7.17] & 0.100 & 0.62 & 0.069 \\
    Attendance & 89.25 (7.90) & 91.63 (8.16) & 2.38 [-3.46, 8.17] & 0.445 & 0.29 & 0.286 \\
    Total & 80.04 (4.12) & 84.91 (3.29) & 4.87 [2.16, 7.48] & 0.003 & 1.29 & 0.009 \\
    \bottomrule
  \end{tabular}

  \vspace{2mm}
  \begin{minipage}{0.95\linewidth}
  \footnotesize
  Note: Hist. = historical control cohort; Int. = intervention cohort; Diff. = intervention minus historical control; \(g\) = Hedges' \(g\); MW = Mann--Whitney \(U\) test. Bracketed intervals are bootstrap 95\% CIs.
  \end{minipage}
\end{table}

Figure~\ref{fig:score_comparison} and Table~\ref{tab:between_cohort_performance} present the score distributions and between-cohort comparisons for the historical control cohort (\(n=12\)) and the intervention cohort (\(n=16\)) across the five rubric dimensions and the weighted total course score. Overall, the intervention cohort showed higher mean scores than the historical control cohort on all assessed dimensions. The mean total course score was 80.04 (SD = 4.12) in the historical control cohort and 84.91 (SD = 3.29) in the intervention cohort, corresponding to a mean difference of 4.87 points with a bootstrap 95\% CI of [2.16, 7.48]. Similar positive mean differences were observed for creativity, design, problem solving, presentation, and attendance.

Welch's two-sample \(t\)-tests indicated statistically significant between-cohort differences for creativity thinking (\(p=0.006\)), design skills (\(p<0.001\)), problem solving (\(p=0.005\)), and total course score (\(p=0.003\)). The corresponding Hedges' \(g\) values were 1.09 for creativity thinking, 1.42 for design skills, 1.23 for problem solving, and 1.29 for total course score, indicating large standardized mean differences in this sample. The bootstrap 95\% CI for these mean differences were also entirely above zero: creativity thinking, 4.88 [1.81, 8.02]; design skills, 5.71 [2.98, 8.35]; problem solving, 6.36 [2.40, 10.00]; and total score, 4.87 [2.16, 7.48]. By contrast, the between-cohort differences in presentation (\(p=0.100\), \(g=0.62\), 95\% CI [-0.29, 7.17]) and attendance (\(p=0.445\), \(g=0.29\), 95\% CI [-3.46, 8.17]) were not statistically significant.

As non-parametric sensitivity analyses, Mann--Whitney \(U\) tests were also conducted for the same between-cohort comparisons. These tests showed a pattern broadly consistent with the Welch's \(t\)-tests, with statistically significant differences for creativity thinking (\(p=0.032\)), design skills (\(p=0.001\)), problem solving (\(p=0.007\)), and total course score (\(p=0.009\)). Presentation showed a non-significant trend in the same direction (\(p=0.069\)), whereas attendance was not statistically significant (\(p=0.286\)). Taken together, the results indicate that the intervention cohort was associated with higher performance most clearly in creativity thinking, design skills, problem solving, and overall course performance.

\subsection{Pre--post changes in questionnaire constructs}

\begin{table}[htbp]
  \centering
  \caption{Pre--post comparison of focal questionnaire constructs in the intervention cohort}
  \label{tab:pre_post_constructs}
  \footnotesize
  \setlength{\tabcolsep}{3pt}
  \begin{tabular}{lcccccc}
    \toprule
    Construct & Pre M (SD) & Post M (SD) & Change [95\% CI] & Pre fav. & Post fav. & Wilcoxon \(p\) \\
    \midrule
    IM & 3.64 (0.55) & 4.16 (0.70) & 0.52 [0.23, 0.83] & 60.9\% & 82.8\% & 0.0072 \\
    SDL & 3.59 (0.61) & 4.08 (0.72) & 0.48 [0.22, 0.75] & 59.4\% & 78.1\% & 0.0058 \\
    \bottomrule
  \end{tabular}

  \vspace{2mm}
  \begin{minipage}{0.95\linewidth}
  \footnotesize
  Note: IM = Innovation and Motivation; SDL = Self-Directed Learning Ability; Fav. = favorable responses, defined as ratings of 4 or 5 on the 5-point Likert scale. Change = post-course minus pre-course construct score. Bracketed intervals are bootstrap 95\% CIs.
  \end{minipage}
\end{table}

\begin{figure}[htbp]
  \centering
  \includegraphics[width=\linewidth]{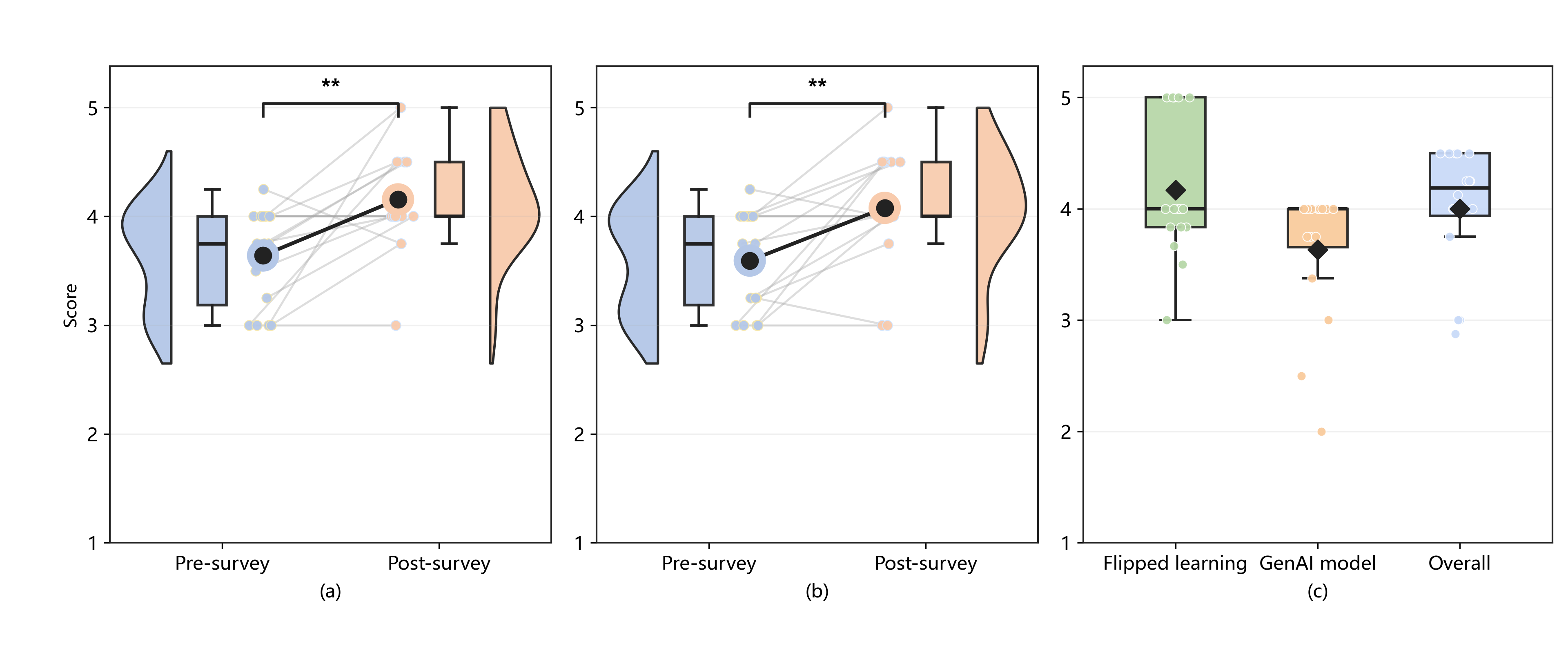}
  \caption{Questionnaire-based comparison of construct scores and post-course evaluations: (a) pre--post change in \textit{Innovation and Motivation}, (b) pre--post change in \textit{Self-Directed Learning Ability}, and (c) post-course ratings for the flipped-learning component, the GenAI module, and the overall learning setting. Asterisks indicate significance levels (* \(p<0.05\), ** \(p<0.01\), *** \(p<0.001\)).}
  \label{fig:survey_comparison}
\end{figure}

Within the intervention cohort, questionnaire responses shifted in a positive direction from the pre-course baseline framing to the post-course experience-based framing (Fig.~\ref{fig:survey_comparison}(a)--(b); Table~\ref{tab:pre_post_constructs}). For \textit{Innovation and Motivation}, the mean construct score increased from 3.64 (SD = 0.55) at pre-test to 4.16 (SD = 0.70) at post-test, corresponding to a mean change of 0.52 with a bootstrap 95\% CI of [0.23, 0.83]. The favorable response rate also increased from 60.9\% to 82.8\%. For \textit{Self-Directed Learning Ability}, the mean construct score increased from 3.59 (SD = 0.61) to 4.08 (SD = 0.72), corresponding to a mean change of 0.48 with a bootstrap 95\% CI of [0.22, 0.75]. The favorable response rate increased from 59.4\% to 78.1\%.

Wilcoxon signed-rank tests indicated statistically significant pre--post differences for both constructs: \textit{Innovation and Motivation} (\(p=0.0072\)) and \textit{Self-Directed Learning Ability} (\(p=0.0058\)). The bootstrap CIs for the mean changes were positive for both constructs, suggesting positive shifts in this sample. Given the non-identical pre- and post-course item framing, these findings are best read as perception-based pre--post comparisons rather than as direct evidence of latent construct growth.

\subsection{Post-course evaluation of the learning setting}

Post-course responses indicated generally positive student perceptions of the learning setting that combined flipped learning and GenAI support (Fig.~\ref{fig:survey_comparison}(c)). At the section level, the flipped-learning component received the highest mean rating (mean = 4.17), followed by the overall class evaluation section (mean = 4.00). The GenAI module received a lower but still positive mean rating (mean = 3.63). These results indicate that students evaluated the overall course design favorably while giving somewhat different ratings to its individual components.

At the item level, the flipped-learning items received consistently high ratings, with mean scores ranging from 4.00 to 4.25. The most highly rated aspects were support for analyzing the relationship between knitwear products and yarn selection and for determining appropriate yarn parameters for specific knitting purposes, both with mean ratings of 4.25. Detailed item-level distributions for these items are reported in Appendix~A, Fig.~\ref{fig:appendix_flipped_class}.

Perceptions of the GenAI module were positive but more varied. Students rated the module highly for expanding creative boundaries and providing ideas they might not otherwise have considered, both with mean ratings of 4.06, and for alignment with creative intentions (mean = 4.00). Lower ratings were observed for saving time in experimentation and preparation (mean = 3.00) and for the extent to which final design outcomes directly incorporated AI-generated references (mean = 2.94). Thus, the GenAI-related ratings were stronger for ideation-oriented items than for time-saving or direct-output-use items. Detailed distributions are presented in Appendix~A, Fig.~\ref{fig:appendix_ai}.

Students also evaluated the overall class setting positively. The highest-rated item concerned effectiveness in achieving learning outcomes (mean = 4.31), while efficiency, interest, ease of understanding, and familiarity all received mean scores above 4.00. Comparatively lower ratings were observed for support for collaboration in portfolio development (mean = 3.62) and for confidence in presenting one's own design work (mean = 3.50), although both ratings remained above the scale midpoint. The corresponding item-level distributions are shown in Appendix~A, Fig.~\ref{fig:appendix_overall}.

\begin{figure}[htbp]
  \centering
  \includegraphics[width=\linewidth]{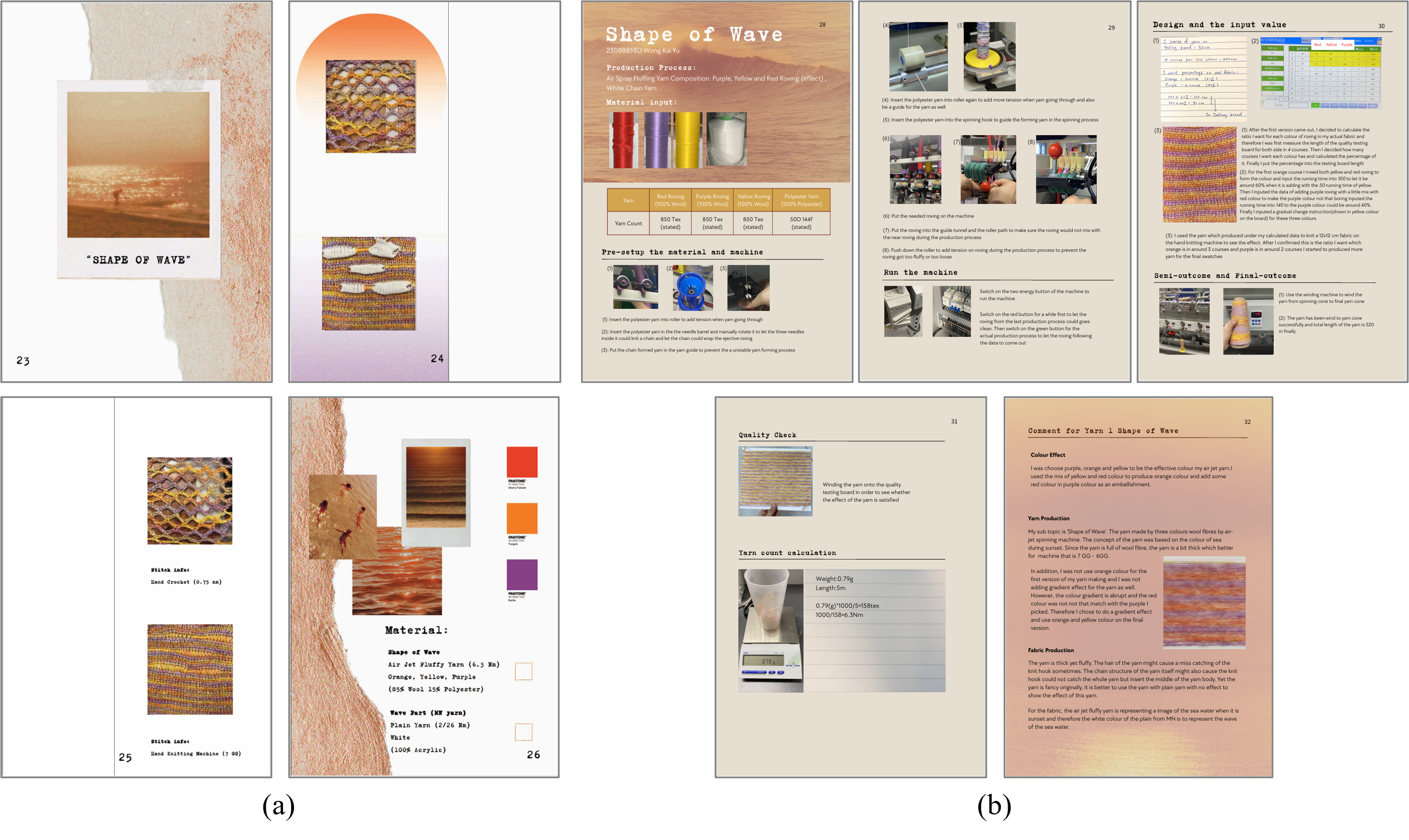}
  \caption{Illustrative example of student problem-solving in the gradient knit swatch project: (a) the final sample included in the portfolio, and (b) the corresponding technical documentation recording the iterative development process.}
  \label{fig:good_example}
\end{figure}

\subsection{Qualitative findings from student and instructor interviews}

To contextualize the quantitative findings, thematic analysis of interviews with four students and the course instructor identified three recurrent themes. Given the small number of interview participants, these findings are used as explanatory context rather than independent evidence of generalizable effects.

First, the flipped-learning arrangement was consistently described as a valuable component of the course. Students reported that the pre-class learning materials helped them understand machine operation, yarn production processes, yarn selection logic, and the sequence of knitting-related procedures before entering the studio. S1 explained that the videos helped the student anticipate difficult parts of the practical work: ``After watching the videos, I had a general idea of the whole process and could identify the parts that might require more attention. So when I was doing the practical work, I was more focused on those difficult steps instead of just following instructions passively.'' The flipped-learning presentation task also appeared to reinforce technical understanding. As S1 further noted, ``When we prepared the PowerPoint presentation, we had to explain the machine and the process to others. That made me review the mechanism more carefully. I think it strengthened my understanding of how the machine works, not only how to operate it.'' Similarly, S2 described the videos as reducing uncertainty before the workshop: ``Because I had already watched the video, I knew the sequence roughly. When I started using the machine, I felt it was easier to get started.'' Overall satisfaction with the course was high, with three students rating the course 5 out of 5 and one student rating it 4 out of 5.

Second, the interviews suggested that students perceived stronger readiness for practical work and broader material understanding. Students described entering studio sessions with clearer procedural knowledge, which they felt reduced the burden associated with first-time equipment use and allowed them to focus more directly on design execution. S2 noted that the videos helped connect pre-class preparation with in-class operation: ``When the teacher explained the machine movement and the order of operation, I could understand it faster because I had already seen it once in the video. I did not need to spend too much time figuring out the basic process.'' Beyond machine operation, S3 emphasized the value of understanding yarn-making and spinning principles for future design practice: ``Before this course, I only had a very general idea about yarn and spinning. But through the videos, the practical work, and the explanation from the teacher, I started to understand the principles behind the process, not just the final product.'' S3 further connected this knowledge to future work, explaining that ``if I understand the production process better, I can design in a more practical and professional way.''

The course instructor similarly observed that students in the intervention cohort appeared to progress more efficiently during the semester. Compared with the prior course offering, students became more comfortable with hands-on work earlier and completed basic sampling sooner. The instructor explained, ``In the past, students usually needed until around Week 11 to complete two samples. This year, many of them were able to finish two samples by around Week 8 or Week 9.'' This earlier progress appeared to create more time for experimentation and refinement. As the instructor noted, ``Because they completed the required work earlier, some students were more willing to try a third sample. This was a positive change, because it meant they had more time to experiment, make mistakes, adjust their process, and improve the final outcome.''

Third, the GenAI module was described as a supplementary tool for early-stage visual ideation. Students used it mainly to explore possible visual directions, while the development and refinement of final design outcomes continued to rely on instructor feedback, material testing, and hands-on studio work. As S4 noted, ``Using the GenAI tool was interesting because it sometimes gave me new ideas, although the generation process was a little slow. I could still use the results as a reference when thinking about my design.''

The instructor interview also highlighted changes in students' portfolio development. In addition to faster progress in swatch production, the instructor observed that students appeared to invest more effort in documenting and presenting their work. The instructor commented, ``Some students paid more attention to the layout, the visual structure, and how to present their process. In previous years, some portfolios were more like a record of finished samples. This year, I saw more evidence of reflection.'' This observation suggests that earlier completion of basic sampling may have created additional space for students to refine not only their physical samples but also the communication and presentation of their design processes.

An illustrative case (Fig.~\ref{fig:good_example}) further highlighted the type of iterative problem-solving observed during the course. In one student's gradient knit swatch project, the initial outcome did not match the intended visual effect because the gradient appearance depended closely on swatch size and related production parameters. Rather than abandoning the idea, the student discussed the issue with the course instructor, re-examined the relevant parameters, recalculated the design plan, and produced a revised sample that better approximated the intended effect. Although anecdotal, this case is consistent with the higher problem-solving scores observed in the intervention cohort and illustrates the kind of iterative reasoning that the course design sought to support.

\section{Discussion}
\label{discussion}

The main contribution of this study is a pedagogical framework for connecting flipped preparation, exemplar reference, GenAI-assisted visual exploration, and studio-based refinement in knit yarn design education. Instead of treating flipped learning and GenAI as separate innovations, the course design assigned them complementary roles: preparing, calibrating, exploring, making, and refining. Pre-class videos and readiness checks supported preparation. The gallery provided feasible examples. The GenAI module supported early visual exploration. Instructor feedback and hands-on studio practice supported feasibility judgement, parameter adjustment, and material realization. This framework helps explain why the clearest outcome differences appeared in creativity thinking, design skills, problem solving, and total course score, while presentation and attendance showed weaker or non-significant differences.

The flipped-learning component appears to have been useful because knit yarn design requires students to coordinate visual intention with yarn properties, colour arrangement, stitch structure, machine operation, scale, and production parameters. A promising idea cannot be judged by appearance alone; it must also be technically feasible. By introducing selected content before class, flipped learning may have reduced uncertainty before studio work and left more contact time for demonstration, guided practice, critique, and iterative making \citep{Shibukawa2019, rasheed2020self, anjomshoaa2022effect, chen2018academic, geng2025}.

The GenAI module contributed in a different and more limited way. It worked mainly as an early-stage visual scaffold, not as a technical advisor. Students used it to explore relationships among colour, yarn type, knitted structure, texture, and overall visual effect before physical sampling. Lower ratings for time saving and direct incorporation into final outcomes show clear limits. Current image-generation systems may not reliably model stitch construction, yarn thickness, tension, scale, or machine constraints. In this course, GenAI was therefore most useful for broadening the early design search space, while technical judgement remained grounded in instructor feedback and studio practice \citep{zhang2024, ray2024, pires2025prompts, Chen2026GenAIMetacognitionDesign}.

The improvement in problem solving can also be read through this structure. The gradient knit swatch case showed how one student moved from a mismatch between intended and physical outcomes to instructor consultation, parameter revision, and a stronger iteration. Although anecdotal, the case reflects the kind of reasoning the course design aimed to support. Prior preparation, visual exploration, exemplar reference, feedback, and repeated sampling may have given students more opportunities to diagnose problems and refine their work, consistent with studio-based learning literature on making, critique, and reflective adjustment \citep{cennamo2011managing,Kumar2021StudioBasedLearning}.

Several limitations remain. The historical-cohort design and absence of random assignment mean that cohort differences or contextual factors may have influenced the results. Although the small cohort size reflects the realities of intensive studio teaching, it limits statistical power and generalizability, especially for the questionnaire and interview findings. Course scores were assigned by a single instructor using a common rubric. The pre- and post-course questionnaire items were also conceptually aligned but not identical repeated measures, so those comparisons should be interpreted as perceived support rather than direct individual growth. Future work should test similar frameworks with larger samples, multiple instructors, and non-historical comparison groups. It should also improve GenAI support for knit-specific structures, yarn parameters, technical documentation, and the translation from visually plausible AI outputs to feasible knitted outcomes.

\section{Conclusions}
\label{conclusions}

This study examined an integrated flipped-learning and GenAI-assisted visual prototyping intervention in an undergraduate knit yarn design course. Across cohorts, the intervention was associated with higher performance in creativity thinking, design skills, problem solving, and total course score. Within the intervention cohort, students also reported positive pre-post shifts in Innovation and Motivation and Self-Directed Learning Ability. Questionnaire and interview findings indicated that flipped learning strengthened readiness for studio work, while GenAI mainly supported early ideation and comparison of visual alternatives. Overall, the findings provide preliminary evidence that structured flipped learning combined with GenAI-assisted visual prototyping can support practice-based studio courses. More broadly, the study offers an instructional framework organized around preparation, calibration, exploration, material realization, and refinement.
\FloatBarrier
\clearpage
\backmatter

\bmhead{Declaration of generative AI and AI-assisted technologies in the writing process}
During the preparation of this work, the authors used ChatGPT/Codex to support language editing, structural revision, and clarity checking. After using this tool, the authors reviewed and edited the content as needed and take full responsibility for the content of the article.

\clearpage

\begin{appendices}

\renewcommand{\theHfigure}{appendix.\thefigure}
\renewcommand{\theHtable}{appendix.\thetable}

\section{Appendix}
\label{app1}

\begin{center}
  \small
  \setlength{\tabcolsep}{6pt}
  \renewcommand{\arraystretch}{1.2}
  
  \begin{longtable}{p{0.08\linewidth} p{0.86\linewidth}}
  \caption{Pre-survey questionnaire items.}
  \label{tab:pre_survey}\\
  
  \multicolumn{2}{p{0.94\linewidth}}{
  \textbf{Instructions:} Please answer the following questions based on your current background and experience. Likert-scale items use a 5-point scale:
  1 = Strongly Disagree, 2 = Disagree, 3 = Neutral, 4 = Agree, 5 = Strongly Agree.
  }\\[6pt]
  
  \hline
  \textbf{No.} & \textbf{Item} \\
  \hline
  \endfirsthead
  
  \multicolumn{2}{r}{\small\emph{(Continued from previous page)}}\\
  \hline
  \textbf{No.} & \textbf{Item} \\
  \hline
  \endhead
  
  \hline
  \multicolumn{2}{r}{\small\emph{(Continued on next page)}}\\
  \endfoot
  
  \hline
  \endlastfoot
  
  \multicolumn{2}{l}{\textbf{Participant information}}\\
  1 & Name and Student No. \\
  2 & Age (Under 18; 18--24; 25--34; 35--44; 45 and above). \\
  3 & Gender (Male; Female; Non-binary; Prefer not to say). \\
  4 & Year of Study (1st year; 2nd year; 3rd year; 4th year or above; Graduate student). \\
  5 & Specialism (open response). \\
  \hline
  
  \multicolumn{2}{l}{\textbf{Yarn knowledge and design/production experience}}\\
  6 & I have a clear understanding of the characteristics of different types of yarn (e.g., material, thickness, usage). \\
  7 & I can easily imagine the final fabric outcome based on the yarn used. \\
  8 & I often find that my lack of familiarity with the production techniques leads to significant differences between my initial design idea and the final product. \\
  9 & I often need extra time to finish assignments related to machine operations (e.g., knitting machines). \\
  \hline
  
  \multicolumn{2}{l}{\textbf{Self-directed learning ability}}\\
  10  & I know what I want to learn. \\
  11 & If there is something I want to learn, I can figure out a way to learn it. \\
  12 & If I discover a need for information that I don't have, I know where to go to get it. \\
  13 & If there is something I have decided to learn, I can find time for it, no matter how busy I am. \\
  \hline
  
  \multicolumn{2}{l}{\textbf{Innovation and motivation}}\\
  14 & I enjoy trying new yarn/fabric-making techniques or design styles. \\
  15 & I feel excited when working on complex yarn/fabric design projects. \\
  16 & I like to challenge myself with difficult yarn/fabric design tasks. \\
  17 & I enjoy researching unfamiliar yarn/fabric-making techniques or technologies. \\
  \hline
  
  \multicolumn{2}{l}{\textbf{AI knowledge and usage}}\\
  18 & In the past month, how often have you used AI to assist your learning or work? (Never; Once or twice; A few times; Several times; Almost daily). \\
  19 & I can give examples of applications where AI assists in design tasks. \\
  20 & What do you use AI for? (Multi-choice: Fashion design image generation; Textile design instruction; Inspiration acquisition; Reports and essay writing; Others [open response]). \\
  \hline
  
  \end{longtable}
  \end{center}

\begin{center}
  \small
  \setlength{\tabcolsep}{6pt}
  \renewcommand{\arraystretch}{1.2}
  
  \begin{longtable}{p{0.08\linewidth} p{0.86\linewidth}}
  \caption{Post-survey questionnaire items.}
  \label{tab:post_survey}\\
  
  \multicolumn{2}{p{0.94\linewidth}}{
  \textbf{Instructions:} Please rate the following statements based on your experience based on the background description. Use a 5-point Likert scale, where:
  1 = Strongly Disagree, 2 = Disagree, 3 = Neutral, 4 = Agree, 5 = Strongly Agree.
  }\\[6pt]
  
  \hline
  \textbf{No.} & \textbf{Item} \\
  \hline
  \endfirsthead
  
  \multicolumn{2}{r}{\small\emph{(Continued from previous page)}}\\
  \hline
  \textbf{No.} & \textbf{Item} \\
  \hline
  \endhead
  
  \hline
  \multicolumn{2}{r}{\small\emph{(Continued on next page)}}\\
  \endfoot
  
  \hline
  \endlastfoot
  
  \multicolumn{2}{l}{\textbf{Participant information}}\\
  1 & Name and Student No. \\
  \hline
  
  \multicolumn{2}{l}{\textbf{Flipped learning}}\\
  \multicolumn{2}{p{0.94\linewidth}}{\emph{In this questionnaire, the flipped learning refers to learning through pre-class videos and materials, followed by in-class presentation and discussions.}}\\
  2 & The flipped learning helped me understand the structure and characteristics of yarn for knitting. \\
  3 & Through the flipped learning, I can describe and produce knitting yarn with different fiber contents and constructions. \\
  4 & The flipped learning helped me analyze the relationship between knitwear products and yarn selection. \\
  5 & The flipped learning helped me determine appropriate yarn parameters (e.g., fiber type, twist level, structure, and weight) for specific knitting purposes. \\
  6 & The flipped learning helped me master the technical operations, thereby reducing the gap between my design and the final product. \\
  7 & The flipped learning helped me complete assignments related to machine operations more efficiently and with less extra time. \\
  \hline
  
  \multicolumn{2}{l}{\textbf{AI agent}}\\
  \multicolumn{2}{p{0.94\linewidth}}{\emph{Based on your learning experience with the support of AI agents, please respond to the following questions.}}\\
  8  & The AI agent expanded my creative boundaries for yarn and fabric design. \\
  9  & The AI agent provided me with design ideas that I wouldn't have considered. \\
  10 & The AI agent helped save my time for experimenting and preparing my design. \\
  11 & The AI-generated designs closely aligned with my initial creative intentions. \\
  12 & The AI agent helped me visualize the relationship between specific yarn characteristics and their resulting fabric effects. \\
  13 & The AI agent helped me better plan the entire process from yarn design to fabric creation. \\
  14 & The AI agent boosted my confidence in exploring new ideas and experimenting with innovative designs. \\
  15 & Did your final design outcome (e.g., yarn or fabric sample) include references from AI-generated results? (1 = Not at all, 5 = Completely) \\
  \hline
  
  \multicolumn{2}{l}{\textbf{Open-ended questions (required)}}\\
  16 & Can you describe a specific instance where the AI agent helped you save time or inspired new ideas in yarn or fabric design? \\
  17 & What challenges did you encounter while using the AI agent for yarn and fabric design, and how did you overcome them? \\
  18 & In your opinion, how can the AI agent be further improved to better assist in yarn and fabric design? \\
  \hline
  
  \multicolumn{2}{l}{\textbf{Overall class setting evaluation}}\\
  \multicolumn{2}{l}{\emph{Comparison with traditional classes}}\\
  19 & The class setting with the flipped learning and AI agent is more effective in achieving learning outcomes. \\
  20 & The class setting with the flipped learning and AI agent is not harder to concentrate on. \\
  21 & The class setting with the flipped learning and AI agent is more efficient. \\
  22 & The class setting with the flipped learning and AI agent is more interesting. \\
  23 & The class setting with the flipped learning and AI agent is easier to understand. \\
  24 & The class setting with the flipped learning and AI agent is more familiar and comfortable. \\
  25 & The class setting with the flipped learning and AI agent helped me collaborate with others to develop and complete a comprehensive portfolio of yarn design projects. \\
  26 & The class setting with the flipped learning and AI agent helped me confidently present and explain my own design work. \\
  \hline
  
  \multicolumn{2}{l}{\textbf{Innovation and motivation}}\\
  27 & The class setting with the flipped learning and AI agent helped me enjoy trying new yarn/fabric-making techniques or design styles. \\
  28 & The class setting with the flipped learning and AI agent helped me feel excited when working on complex yarn/fabric-design projects. \\
  29 & The class setting with the flipped learning and AI agent helped me like to challenge myself with difficult yarn/fabric-design tasks. \\
  30 & The class setting with the flipped learning and AI agent helped me enjoy researching unfamiliar yarn/fabric-making techniques or technologies. \\
  \hline
  
  \multicolumn{2}{l}{\textbf{Self-directed learning ability}}\\
  31 & The class setting with the flipped learning and AI agent helped me know what I want to learn. \\
  32 & The class setting with the flipped learning and AI agent helped me figure out how to learn something when I want to learn it. \\
  33 & The class setting with the flipped learning and AI agent helped me know where to find information when I realized I need it. \\
  34 & The class setting with the flipped learning and AI agent helped me manage my time to learn new things, even when I was busy. \\
  
  \end{longtable}
  \end{center}

\begin{figure}[h]
  \centering
  \includegraphics[width=\linewidth,height=0.86\textheight,keepaspectratio]{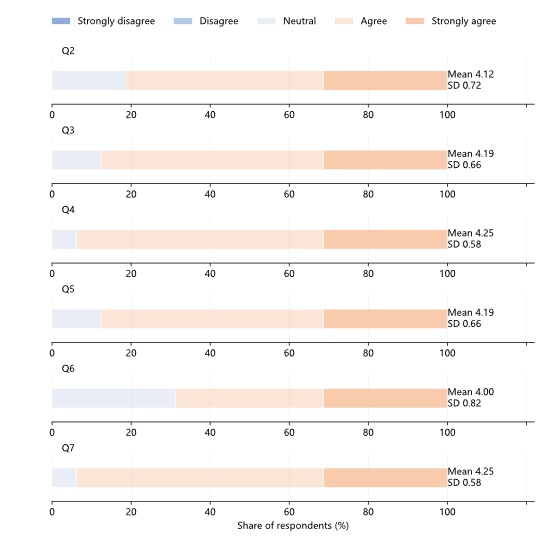}
  \caption{Post-survey item-level Likert distributions for perceived effects of the flipped learning.}
  \label{fig:appendix_flipped_class}
\end{figure}

\begin{figure}[h]
  \centering
  \includegraphics[width=\linewidth,height=0.95\textheight,keepaspectratio]{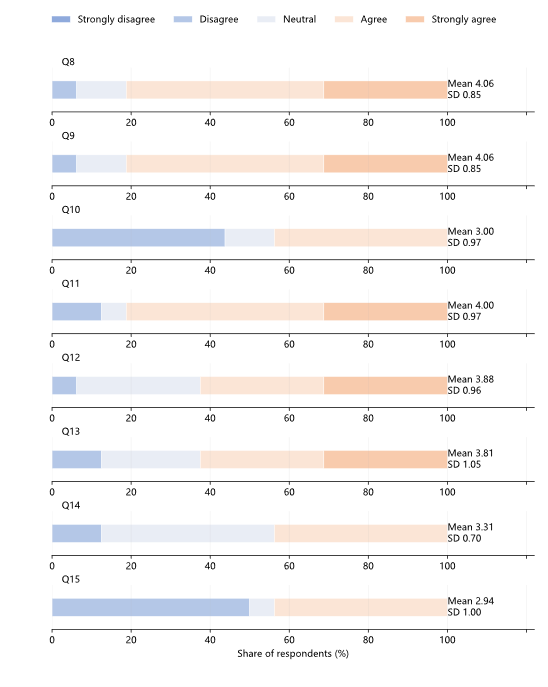}
  \caption{Post-survey item-level Likert distributions for perceived effects of the AI agent.}
  \label{fig:appendix_ai}
\end{figure}

\begin{figure}[h]
  \centering
  \includegraphics[width=\linewidth,height=0.95\textheight,keepaspectratio]{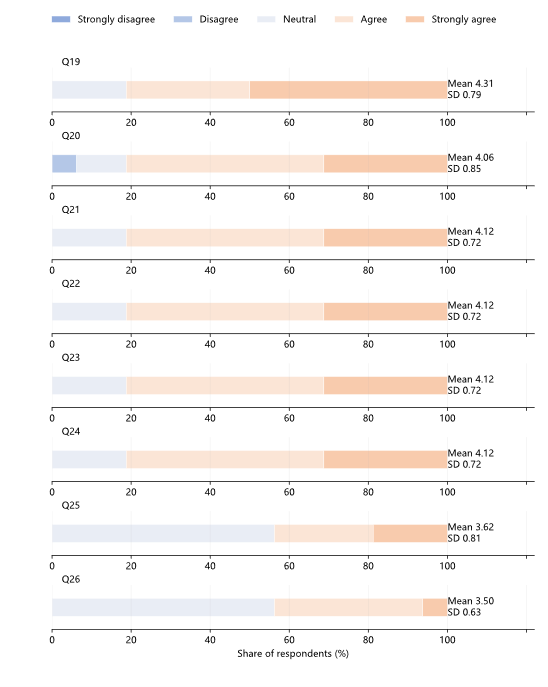}
  \caption{Post-survey item-level Likert distributions for overall evaluation of the flipped-classroom-and-AI-agent learning setting.}
  \label{fig:appendix_overall}
\end{figure}

\end{appendices}

\end{document}